\documentclass{elsart}

\usepackage[square,comma]{natbib}
\usepackage{graphicx}
\usepackage{pxfonts}
\usepackage{lineno}

\usepackage{amssymb}

\journal{}

\begin{document}

\thispagestyle{empty}
\begin{Large}
\textbf{DEUTSCHES ELEKTRONEN-SYNCHROTRON}

\textbf{\large{Ein Forschungszentrum der Helmholtz-Gemeinschaft}\\}
\end{Large}

DESY 12-051

March 2012

\begin{eqnarray}
\nonumber &&\cr \nonumber && \cr \nonumber &&\cr
\end{eqnarray}
\begin{eqnarray}
\nonumber
\end{eqnarray}
\begin{center}
\begin{Large}
\textbf{Pulse-front tilt caused by the use of a grating
monochromator and self-seeding of soft X-ray FELs}
\end{Large}
\begin{eqnarray}
\nonumber &&\cr \nonumber && \cr
\end{eqnarray}

\begin{large}
Gianluca Geloni,
\end{large}
\textsl{\\European XFEL GmbH, Hamburg}
\begin{large}

Vitali Kocharyan and Evgeni Saldin
\end{large}
\textsl{\\Deutsches Elektronen-Synchrotron DESY, Hamburg}
\begin{eqnarray}
\nonumber
\end{eqnarray}
\begin{eqnarray}
\nonumber
\end{eqnarray}
ISSN 0418-9833
\begin{eqnarray}
\nonumber
\end{eqnarray}
\begin{large}
\textbf{NOTKESTRASSE 85 - 22607 HAMBURG}
\end{large}
\end{center}
\clearpage
\newpage

\begin{frontmatter}



\title{Pulse-front tilt caused by the use of a grating monochromator and self-seeding of soft X-ray FELs}


\author[XFEL]{Gianluca Geloni\thanksref{corr},}
\thanks[corr]{Corresponding Author. E-mail address: gianluca.geloni@xfel.eu}
\author[DESY]{Vitali Kocharyan}
\author[DESY]{and Evgeni Saldin}

\address[XFEL]{European XFEL GmbH, Hamburg, Germany}
\address[DESY]{Deutsches Elektronen-Synchrotron (DESY), Hamburg,
Germany}

\begin{abstract}
Self-seeding is a promising approach to significantly narrow the
SASE bandwidth of XFELs to produce nearly transform-limited pulses.
The development of such schemes in the soft X-ray wavelength range
necessarily involves gratings as dispersive elements. These
introduce, in general, a pulse-front tilt, which is directly
proportional to the angular dispersion.  Pulse-front tilt may easily
lead to a seed signal decrease by a factor two or more. Suggestions
on how to minimize the pulse-front tilt effect in the self-seeding
setup are given.
\end{abstract}

%
%
%
\end{frontmatter}



\section{\label{sec:intro} Introduction}

As a consequence of the start-up from shot noise, the longitudinal
coherence of  X-ray SASE FELs is rather poor compared to
conventional optical lasers. Self-seeding schemes have been studied
to reduce the bandwidth of SASE X-ray FELs \cite{SELF}-\cite{OURS}.
In general, a self-seeding setup consists of two undulators
separated by a photon monochromator and an electron bypass, normally
a four-dipole chicane. For soft X-ray self-seeding , a monochromator
usually consists of a grating \cite{SELF}. Recently, a very compact
soft X-ray self-seeding scheme has appeared, based on grating
monochromator \cite{FENG,FENG2}.

In \cite{OURS} we studied the performance of this compact scheme for
the European XFEL upgrade. Limitations on the performance of the
self-seeding scheme related with aberrations and spatial quality of
the seed beam have been extensively discussed in \cite{FENG,FENG2}
and go beyond the scope of this paper. Here we will focus our
attention on the spatiotemporal distortions of the X-ray seed pulse.
Numerical results provided by ray-tracing algorithms applied to
grating design programs give accurate information on the spatial
properties of the imaging optical system of grating monochromator.
However, in the case of self-seeding, the spatiotemporal deformation
of the seeded X-ray optical pulses is not negligible: aside from the
conventional aberrations, distortions as pulse-front tilt should
also be considered \cite{WEIN,HEBL,PRET}. The propagation and
distortion of X-ray pulses in grating monochromators can be
described using a wave optical theory. Most of our calculations are,
in principle, straightforward applications of conventional ultrafast
pulse optics \cite{WEIN}. Our paper provides physical understanding
of the self-seeding setup with a grating monochromator, and we
expect that this study can be useful in the design stage of
self-seeding setups.

\section{\label{sec:teo} Theoretical background for the analysis of pulse-front tilt
phenomena}

\subsection{Pulse-front tilt from gratings}

\begin{figure}[tb]
\includegraphics[width=0.5\textwidth]{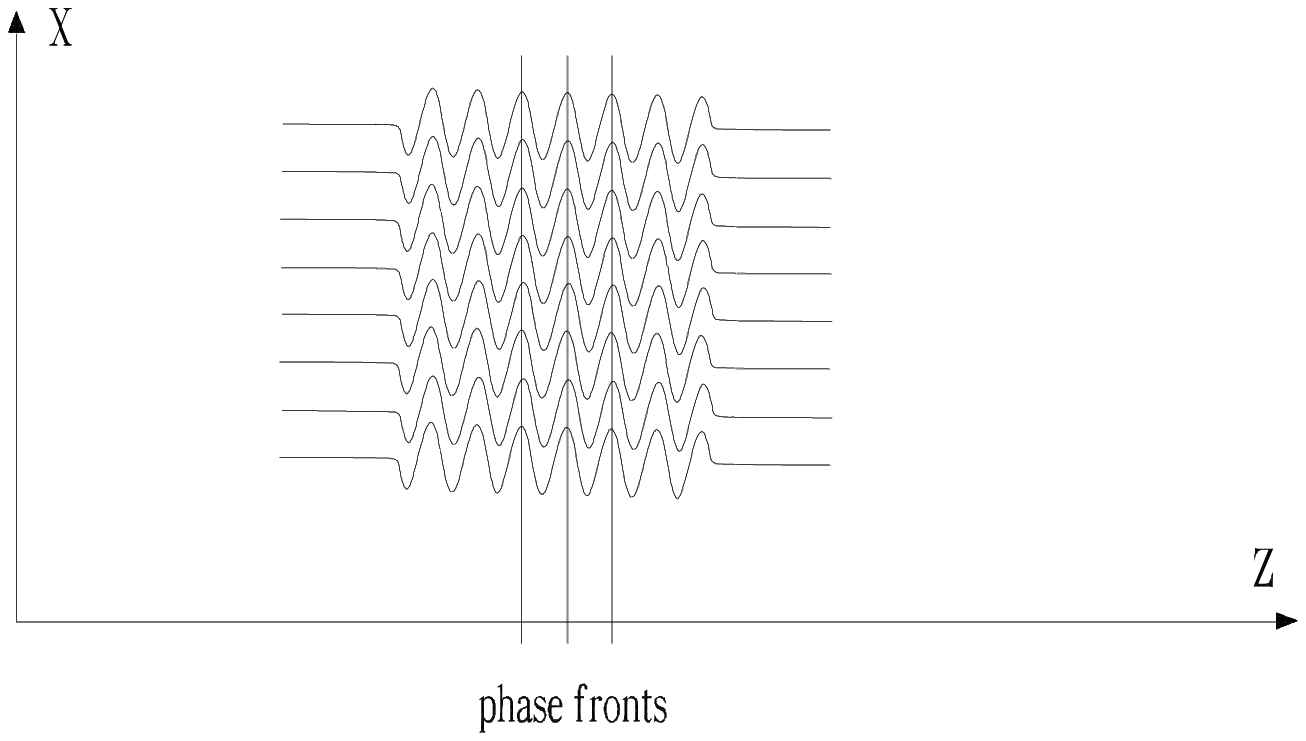}
\includegraphics[width=0.5\textwidth]{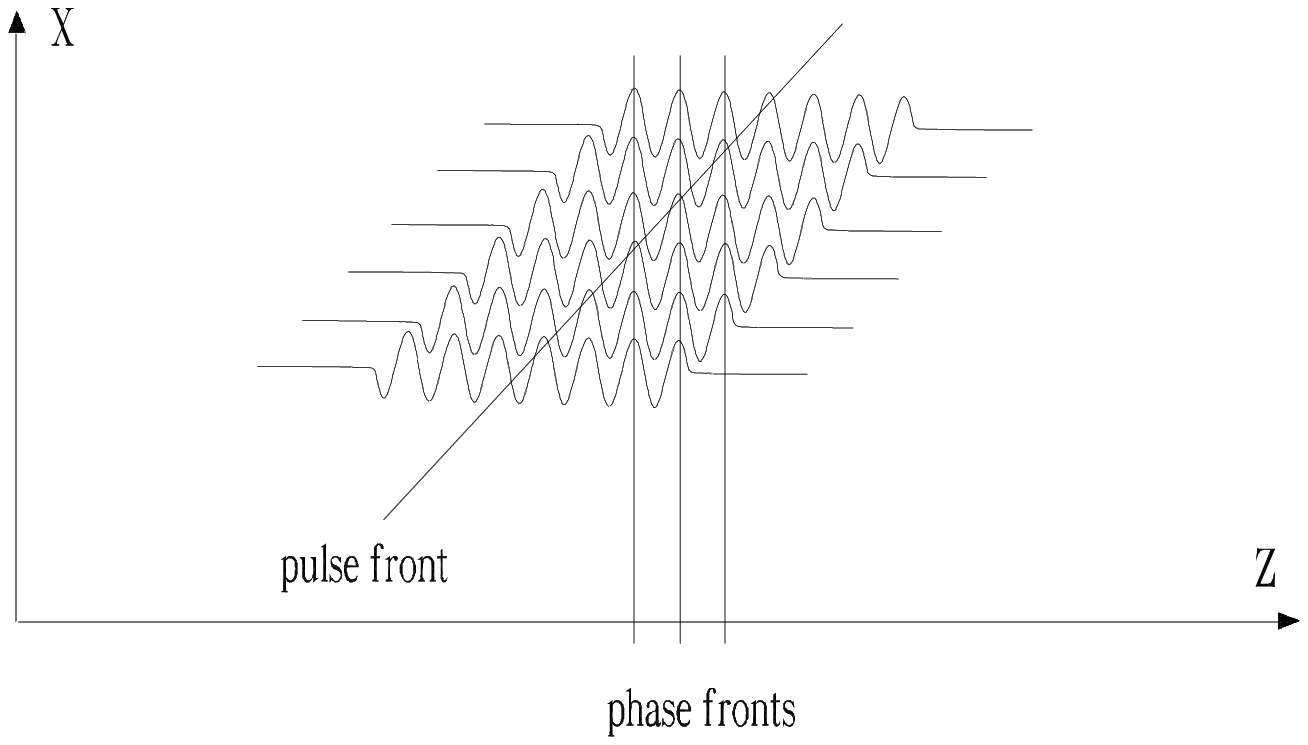}
\caption{Schematic representation of the electric field profile of
an undistorted pulse beam  (left) and of a beam with pulse front
tilt (right). The z axis is  along the beam propagation direction
(adapted from \cite{PRET}).} \label{grat1}
\end{figure}

\begin{figure}[tb]
\includegraphics[width=1.0\textwidth]{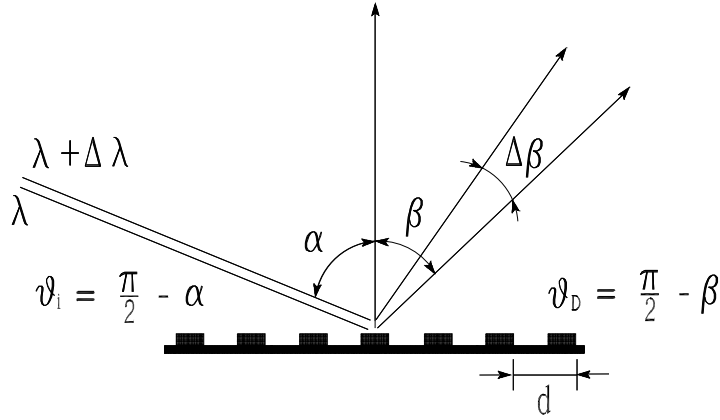}
\caption{Geometry of diffraction grating scattering.} \label{grat2}
\end{figure}

Ultrashort X-ray FEL pulses are usually represented as products of
electric field factors separately dependent on space and time. The
assumption of separability of the spatial (or spatial frequency)
dependence of the pulse from the temporal (or temporal frequency)
dependence is usually made for the sake of simplicity. However, when
the manipulation of ultrashort X-ray pulses requires the
introduction of coupling between spatial and temporal frequency
coordinates, such assumption fails. The direction of energy flow
-usually identified as rays directions- is always orthogonal to the
surface of constant phase, that is to the wavefronts of the
corresponding propagating wave. If one deals with ultrashort X-ray
pulses, one has to consider, in addition, planes of constant
intensity, that is pulse fronts. Fig. 1 shows a schematic
representation of the electric field profile of an undistorted pulse
and one with a pulse-front tilt. A distortion of the pulse front
does not affect propagation, because the phase fronts remain
unaffected. However, for most applications, including self-seeding
applications, it is desirable that these fronts be parallel to the
phase fronts, and therefore orthogonal to the propagating direction.

A pulse-front tilt can be present in the beam due to the propagation
through a grating monochromator. As shown in Fig. 2, the input beam
is incident on the grating at an angle $\theta_i$. The diffracted
angle $\theta_D$ is a function of frequency, according to the
well-known plane grating equation. Assuming diffraction into the
first order, one has

\begin{eqnarray}
\lambda = d (\cos \theta_i - \cos \theta_D)~, \label{grateq}
\end{eqnarray}
where $\lambda = 2 \pi c/\omega$, and $d$ is the groove spacing. Eq.
(\ref{grateq}) describes the basic working of a grating
monochromator. By differentiating this equation one obtains

\begin{eqnarray}
\frac{d \theta_D}{d \lambda} = \frac{1}{\theta_D d} ~,
\label{diffgrat}
\end{eqnarray}
where we assume grazing incidence geometry, $\theta_i \ll 1$ and
$\theta_D \ll 1$. The physical meaning of Eq. (\ref{diffgrat}) is
that different spectral components of the outcoming pulse travel in
different directions. The electric field of a pulse including
angular dispersion can be expressed in the Fourier domain $\{k_x,
\omega\}$ as $E(k_x - p \omega, \omega)$, while the inverse Fourier
transform from the $\{k_x, \omega\}$ domain to the space-time domain
$\{x,t\}$ can be expressed as $E(x, t + p x)$, which is the electric
field of a pulse with a pulse-front tilt. The tilt angle $\gamma$ is
given by $\tan \gamma = c p$. More specifically

\begin{eqnarray}
p = \frac{d k_x}{d \omega} = k \frac{d \theta_D}{d \omega} =
\frac{\lambda}{c} \frac{d \theta_D}{d \lambda} = \frac{\lambda}{c
\theta_D d}~. \label{pgiven}
\end{eqnarray}
Therefore one concludes that the pulse-front tilt is invariably
accompanied by angular dispersion. It follows that any device like a
grating monochromator, that introduces an angular dispersion, also
introduces significant pulse-front tile, which is problematic for
seeding.

\subsection{Spatiotemporal transformation of X-ray FEL  pulses by
crystals}

The development of  self-seeding schemes in the hard X-ray
wavelength range necessarily involves crystal monochromators.
Recently, the spatiotemporal coupling in the electric field relevant
to self-seeding schemes with crystal monochromators has been
analyzed in the frame of classical dynamical theory of X-ray
diffraction \cite{SHVID}. This analysis shows that a crystal in
Bragg reflection geometry transforms the incident electric field
$E(x,t)$ in the $\{x,t\}$ domain into $E(x- a t, t)$, that is the
field of a pulse with a less well-known distortion, first studied in
\cite{GABO}. The physical meaning of this distortion is that the
beam spot size is independent of time, but the beam central position
changes as the pulse evolves in time. One of the aims of this
subsection is to disentangle what is specific to the transformation
by a crystal and what is intrinsic to the grating case. Our purpose
here is not that of presenting novel results but, rather, to attempt
a more intuitive explanation of spatiotemporal coupling phenomena in
the dynamical theory of X-ray diffraction, and to convey the
importance and simplicity of the results presented in \cite{SHVID}.

\begin{figure}[tb]
\includegraphics[width=1.0\textwidth]{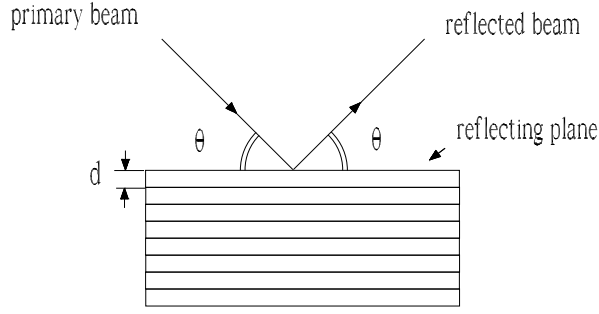}
\caption{Reflection of the primary beam from the lattice planes in
the crystal according to Bragg law.} \label{grat3}
\end{figure}
We begin our analysis by specifying the scattering geometry under
study. The angle between the physical surface of the crystal and the
reflecting atomic planes is an important factor. The reflection is
said to be symmetric if the surface normal is perpendicular to the
reflecting planes in the case of Bragg geometry. We shell examine
only the symmetric Bragg case, Fig. \ref{grat3}.

Let us consider an electromagnetic plane wave in the X-ray frequency
range incident on an infinite, perfect crystal. Within the
kinematical approximation, according to the Bragg law, constructive
interference of waves scattered from the crystal occurs if the angle
of incidence, $\theta_i$ and the wavelength, $\lambda$, are related
by the well-known relation

\begin{eqnarray}
\lambda = 2 d \sin \theta_i ~. \label{bragg}
\end{eqnarray}
assuming reflection into  the first order. This equation shows that
for a given wavelength of the X-ray beam, diffraction is possible
only at certain angles determined by the interplanar spacings $d$.
It is important to remember the following geometrical relationships:

1. The angle between the incident X-ray beam and normal to the
reflection plane is equal to that between the normal and the
diffracted X-ray beam. In other words, Bragg reflection is a mirror
reflection, and the incident angle is equal to the diffracted angle
($\theta_i = \theta _D$).

2. The angle between the diffracted X-ray beam and the transmitted
one is always $2 \theta_i$. In other words, incident beam and
forward diffracted (i.e. transmitted) beam have the same direction.

We now turn our attention beyond the kinematical approximation to
the dynamical theory of X-ray diffraction by a crystal. An optical
element inserted into the X-ray beam is supposed to modify some
properties of the beam as its width, its divergence, or its spectral
bandwidth. It is useful to describe the modification of the beam by
means of a transfer function. The reflectivity curve - the
reflectance - in Bragg geometry can be expressed in the frame of
dynamical theory as

\begin{eqnarray}
R(\theta_i,\omega) = R(\Delta \omega +  \omega_B \Delta \theta \cot
\theta_B  ) ~, \label{reflectance}
\end{eqnarray}
where $\Delta \omega = (\omega - \omega_B)$ and $\Delta \theta =
(\theta_i - \theta_B)$ are the deviations of frequency and incident
angle of the incoming beam from the Bragg frequency and Bragg angle,
respectively. The frequency $\omega_B$ and the angle $\theta_B$ are
given by the Bragg law: $\omega_B \sin \theta_B = \pi c/d$. We
follow the usual procedure of expanding $\omega$ in a Taylor series
about $\omega_B$, so that

\begin{eqnarray}
\omega = \omega_B + (d \omega/d \theta)_B (\theta - \theta_B) + ...
~. \label{taylor}
\end{eqnarray}
Consider a perfectly collimated, white beam incident on the crystal.
In kinematical approximation $R$ is a Dirac $\delta$-function, which
is simply represented by the differential form of Bragg law:

\begin{eqnarray}
d \lambda/d \theta_i = \lambda \cot \theta_i  ~. \label{difform}
\end{eqnarray}
In contrast to this, in dynamical theory the reflectivity width is
finite. This means that there is a reflected beam even when incident
angle and wavelength of the incoming beam are not related exactly by
Bragg equation. It is interesting to note that the geometrical
relationships 1. and 2. are still valid in the framework of
dynamical theory. In particular, reflection in dynamical theory is
always a mirror reflection. We underline here that if we have a
perfectly collimated, white incident beam, we also have a perfectly
collimated reflected beam. Its bandwidth is related with the width
of the reflectivity curve. We will regard the beam as perfectly
collimated when the angular spread of the beam is much smaller than
the angular width of the transfer function $R$. It should be
realized that the crystal does not introduce an angular dispersion
similar to a grating or a prism. However, a more detailed analysis
based on the expression for the reflectivity, Eq.
(\ref{reflectance}), shows that a less well-known spatiotemporal
coupling exists. The fact that the reflectivity is invariant under
angle and frequency transformations obeying

\begin{eqnarray}
\Delta \omega +  \omega_B \Delta \theta \cot \theta_B  =
\mathrm{const}~ \label{transform}
\end{eqnarray}
is evident, and corresponds to the coupling in the Fourier domain
$\{k_x, \omega\}$. The origin of this relation is kinematical, it is
due to Bragg diffraction. One might be surprised that the field
transformation derived in \cite{SHVID} for an XFEL pulse after a
crystal in the $\{x,t\}$ domain is given by

\begin{eqnarray}
E_{\mathrm{out}} (x,t) =  FT[R(\Delta \omega,
k_x)E_{\mathrm{in}}(\Delta \omega, k_x)] = E(x- c t\cot \theta_B ,
t)~, \label{Eoutxt}
\end{eqnarray}
where $FT$ indicates a Fourier transform from the $\{k_x,\omega\}$
to the $\{x,t\}$ domain, and $k_x = \omega_B \Delta \theta/c$. In
general, one would indeed expect the transformation to be symmetric
in both the $\{k_x,\omega\}$ and in the $\{x,t\}$ domain due to the
symmetry of the transfer function\footnote{There is a breaking of
the symmetry in the diffracted beam in the $\{k_x,\omega\}$ domain.
While the symmetry is present at the level of the transfer function,
it is not present anymore when one considers the incident beam. We
point out that symmetry breaking in \cite{SHVID} is a result of the
approximation of temporal profile of the incident wave to a Dirac
$\delta$-function.}. However, it is reasonable to expect the
influence of a nonsymmetric input beam distribution. In the
self-seeding case, the incoming XFEL beam is well collimated,
meaning that its angular spread is  a few times smaller than angular
width of the transfer function\footnote{In \cite{OURY4} we pointed
out that: "In our case of interest (hard X-ray self-seeding with
wake monochromator) we have an angular divergence of the incident
photon beam of about a microradian, which is much smaller than the
Darvin's width of the rocking curve (10 microradians). As a result,
we assume that all frequencies impinge on the crystal at the same
angle in the vicinity of the Bragg diffraction condition. Note that
mirror reflection takes place for all frequencies and, therefore,
the reflected beam has exactly the same divergence as the incoming
beam". In other words, the description of our problem includes a
small parameter, the small ratio between the beam angular width and
the width of the crystal transfer function. This justifies the
application in \cite{OURY4} (with an accuracy of about $10 \%$) of
the plane wave approximation for the first transmission maximum. We
thus avoided difficulties related with spatiotemporal coupling.}.
Only the bandwidth of the incoming beam is much wider than the
bandwidth of the transfer function. In this limit, we can
approximate the transfer function in the expression for the inverse
temporal Fourier transform as a Dirac $\delta$-function. This gives

\begin{eqnarray}
&& E_\mathrm{out}(x,t) = FT[R(\Delta \omega, k_x) E_\mathrm{in}
(\Delta \omega, k_x)]\cr && = \xi(t) \cdot \frac{1}{2\pi}  \int d
k_x \exp(-i k_x c t \cot \theta_B) \exp(i k_x x)
E_\mathrm{in}(0,k_x)\cr && = \xi(t) b(x - c t\cot \theta_B) ~,
\label{eoutdel}
\end{eqnarray}
where we applied the Shift Theorem twice, and where

\begin{eqnarray}
\xi(W) = \frac{1}{2\pi} \int d Y \exp( i Y W) R(Y) \label{tempFT}
\end{eqnarray}
is the inverse Fourier transform of the reflectivity curve.

In the opposite limit when the incoming beam has a wide angular
width and a narrow bandwidth we take the transfer function in the
inverse spatial Fourier transform as a Dirac $\delta$-function. This
gives

\begin{eqnarray}
E_\mathrm{out}(x,t) = \xi(x \tan \theta_B/c)  a(t - (x/c) \tan
\theta_B )~ , \label{Eoutother}
\end{eqnarray}
where $\xi(x)$ is given in Eq. (\ref{tempFT}). These two limits
represent the two sides of the symmetry of the transfer function.

The last expression, Eq. (\ref{Eoutother}) is the field of a pulse
with a pulse front tilt. Typically one would think that a pulse
front tilt can be introduced only by dispersive elements like
gratings or prisms. Here we presented an example in which no
dispersive elements exists, and we stress that angular dispersion
can be introduced by non dispersive element like crystals too.

Although we began by considering a case of reflection transfer
function in Bragg reflection geometry, none of our arguments depends
on that fact. Eq. (\ref{reflectance}) still holds if the transfer
function R is referred to the transmittance in Bragg reflection
geometry. For the transmitted beam, all derivations are worked out
in the same way we have done here and gives asymptotic expression
like Eq(\ref{eoutdel}) , Eq. (\ref{Eoutother}) for field of forward
scattered pulses.

\section{\label{sec:model} Modeling of self-seeding setup with grating
monochromator}

\begin{figure}[tb]
\includegraphics[width=1.0\textwidth]{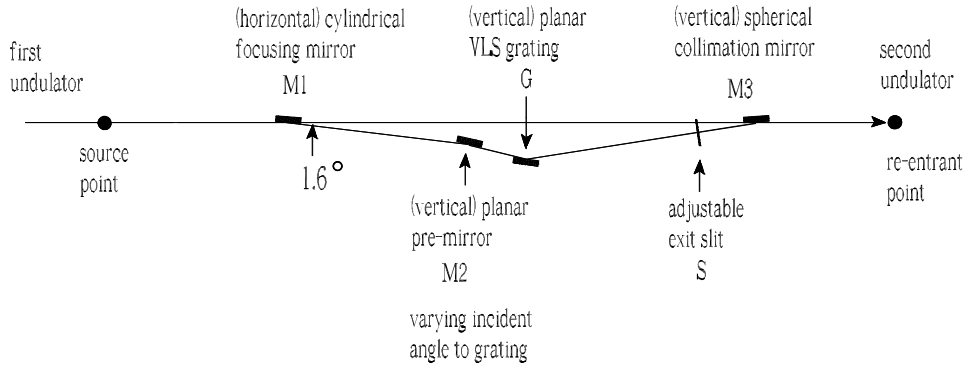}
\caption{Optics for the compact grating monochromator originally
proposed at SLAC \cite{FENG,FENG2} for the soft X-ray self-seeding
setup. } \label{grat4}
\end{figure}

A self-seeding setup should be compact enough to fit one undulator
segment. In this case its installation does not perturb the
undulator focusing system and allows for the safe return to the
baseline mode of operation. The design adopted for the LCLS  is the
novel one by Y. Feng et al. \cite{FENG,FENG2}, and is based on a
planar VLS grating. It is equipped only with an exit slit. Such
design includes four optical elements, a cylindrical and spherical
focusing mirrors, a VLS grating and a plane mirror in front of the
grating. The optical layout of the monochromator is schematically
shown in Fig. \ref{grat4}.

\begin{figure}[tb]
\includegraphics[width=1.0\textwidth]{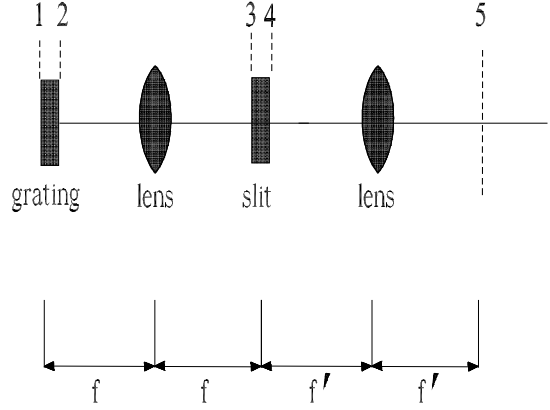}
\caption{Diagram of the self-seeding grating monochromator used in
theoretical analysis. For simplicity the grating is depicted in
transmission mode, and  tilt is not shown. The dashed lines refer to
different planes at which the field is calculated (adapted from
\cite{WEIN}).} \label{grat5}
\end{figure}
A simplified diagram for analyzing the grating monochromator is
shown in Fig. \ref{grat5}. We will assume that the optical system
used for imaging purposes is the well-known two-lens image formation
system. With reference to Fig. \ref{grat4}, the VLS grating is
represented by a combination of a planar grating with fixed line
spacing and a lens, with the focal length of the lens equal to the
focal length of the VLS grating. The analysis of the grating
monochromator is simplified by recognizing that the grating can be
shifted from a position immediately before the lens to a position
immediately after the object plane. The monochromator is treated
assuming no aberrations. This approximation is useful, since for the
design shown on Fig. \ref{grat4} the aberration effects are
negligible \cite{FENG,FENG2}. This simplifies calculations and
allows analytical results to be derived.

The angular dispersion of the grating causes a separation of
different optical frequencies at the Fourier plane of the first
focusing element (lens). Therefore, this system becomes a tunable
frequency filter if a slit is placed at the Fourier plane. We assume
that the two lenses in Fig. \ref{grat5} are not identical, so that
this scheme allows for magnification by changing the focal distance
of the second lens.

It is important to analyze the output field from the grating
monochromator quantitatively. In our analysis we calculate the
propagation of the input signal to different planes of interest
within the self-seeding monochromator, as indicated in Fig.
\ref{grat5}. We start by writing the input field in plane 1,
immediately before the grating, as

\begin{eqnarray}
E_1(x,t) = \mathrm{Re} [a(t)b(x) \exp(i \omega_0 t)]~,
\label{beforegrating}
\end{eqnarray}
where $\omega_0$ is the pulse carrier frequency, which is linked to
$k$ by $k = \omega_0/c$. We assume that the input signal is Gaussian
in the transverse direction, that is $b(x) = 1/\left(\sqrt{2\pi}
\sigma\right) \exp(- x^2/2 \sigma^2)$. The field in plane 2,
immediately after the grating (assuming diffraction into the first
order) may be written as

\begin{eqnarray}
E_2(x,t) = \frac{\sqrt{\beta}}{2\pi} \mathrm{Re} \int d \Delta
\omega A(\Delta \omega) b(\beta x) \exp(i p \Delta \omega x) \exp[i
\omega t], \label{aftergrating}
\end{eqnarray}
where the astigmatism factor $\beta = \theta_i/\theta_D$, results
due to the difference in input and output angles, and $p =
\lambda/(c\theta_D d)$.

Our analysis exploits the Fourier transform properties of a lens. In
particular we consider the propagation of a monochromatic
one-dimensional field in paraxial approximation.  The field
distribution in the focal plane of the first lens, which we call
plane 3, is given by

\begin{eqnarray}
E_3(x,t) = \frac{1}{2 \pi \sqrt{ \lambda f \beta}} \mathrm{Re} \int
d \Delta \omega \sqrt{i} A(\Delta \omega) \hat{b}\left(
\frac{k}{\beta f}(x+\eta \Delta \omega)\right)\exp(i \omega t)~,
\label{plane5}
\end{eqnarray}
where $\hat{b}(k_x)$ refers to the spatial Fourier transform of the
input spatial profile, $\eta = f \lambda^2/(2\pi c d\theta_D)$ is a
spatial dispersion parameter, which describes the proportionality
between spatial displacement and optical frequency. In the case of a
Gaussian input beam we have $\hat{b}(k_x) = \exp(- \sigma^2
k_x^2/2)$. Therefore, the field in the Fourier plane is written as

\begin{eqnarray}
E_3(x,t) =  \frac{1}{2 \pi \sqrt{ \lambda f \beta}} \mathrm{Re} \int
d \Delta \omega \sqrt{i} A(\Delta \omega) \exp\left[-\frac{(x+\eta
\Delta \omega)^2}{2 \sigma_f^2}\right]\exp(i \omega t)~, \cr &&
\label{plane3}
\end{eqnarray}
where $\sigma_f = \beta f/(k \sigma)$ is the rms of the focused beam
at the Fourier plane for any single frequency component.

We now add a slit at the Fourier plane, that we regard as a
particular spatial mask with a real transmission function $S(x)$.
The field in plane 4, that is directly after the slit is simply
given by

\begin{eqnarray}
E_4(x,t) = S(x) E_3(x,t)~. \label{plane4}
\end{eqnarray}
Let us first consider the limiting case of a $\delta$-function slit
that is, physically, a slit with much narrower opening than the spot
size of a fixed individual frequency, centered at transverse
position $x'$. For a Gaussian input beam, the square modulus of the
transmittance of the monochromator, that is the frequency response,
is given by

\begin{eqnarray}
|T(\Delta \omega)|^2 = A_0 \exp\left[ - \left(\frac{\lambda
\sigma}{c d \theta_i}\right)^2 (\Delta \omega - \Delta
\omega')^2\right] ~.
\end{eqnarray}
The center frequency of the passband , $\Delta \omega' = x'/\eta$,
is determined by the transverse position of the slit.  The spectral
resolution of the monochromator depends on the spot size at the
Fourier plane related with the individual frequencies, $\sigma_f$,
and on the rate of spatial dispersion with respect to the frequency
(determined by $\eta$). The FWHM of the monochromator spectral line
is

\begin{eqnarray}
\frac{\Delta \omega}{\omega} =  1.18 \frac{d \theta_i}{2 \pi \sigma}
\sim \frac{1}{\pi N}~, \label{Dww}
\end{eqnarray}
where $N \sim 2 \sigma/(d \theta_i)$ is the number of grooves
illuminated by the input beam. For any single frequency the spot
size at the Fourier plane, and hence the bandwidth transmitted
through a narrow slit, is inversely proportional to the input spot
size.  Since the temporal spread in the output pulse is inversely
proportional to the transmitted bandwidth, the output pulse duration
is proportional to the input spot size.

To get to the plane prior to the second undulator entrance, that
will be called plane 5, we simply perform a second spatial Fourier
transform. The resulting field is located at an output plane at
distance $f'$ behind the second lens with focal distance $f'$, and
is given by

\begin{eqnarray}
&&E_5(x,t) = \frac{1}{2 \pi \lambda  \sqrt{f' f \beta}} \mathrm{Re}
\int d \Delta \omega ~i A(\Delta \omega) \exp(i \omega t) \cr &&
\times \int \int d x' d x'' S(x'') b(x') \exp\left[\frac{i k
x'(x''+\eta \Delta \omega)}{\beta f}\right] \exp\left(\frac{i k x
x''}{f'}\right) \label{plane5a}
\end{eqnarray}

Performing the integral over $x''$ first we obtain

\begin{eqnarray}
&& E_5(x,t) = \frac{1}{4\pi^2} \sqrt{\frac{f \beta}{f'}} \mathrm{Re}
\int d \Delta \omega ~i A(\Delta \omega) \exp(i \omega t) \cr &&
\times \int d X \exp(i \eta \Delta \omega X) b(\beta f X/k)
\hat{S}(X + k x/f ') ~,\label{plane5b}
\end{eqnarray}
where we define

\begin{eqnarray}
\hat{S}(k) = \int dx S(x) \exp[i k x] ~.\label{Sk}
\end{eqnarray}
We now turn to consider the limiting case when the spot size of any
given individual frequency at the Fourier plane is small compared to
the spatial scale over which the transmission of the slits varies.
This limiting situation is the opposite of the previously analyzed
case of $\delta$-function slits; here the slits do not modify the
spatial profiles of the individual frequency components.
Mathematically this corresponds to the substitution of the function
$\hat{S}$ in the expression for $E_5(x,t)$ with a Dirac
$\delta$-function. In other words we assume that slits are absent.
One obtains

\begin{eqnarray}
E_5(x,t) = \mathrm{Re}\Bigg[&&\frac{i}{2\pi} \sqrt{\frac{f \beta}{f
'}} b\left(- \frac{\beta f x}{f'}\right) \cr && \times \int d \Delta
\omega A(\Delta \omega) \exp\left(\frac{-i p \Delta \omega f x
}{f'}\right) \exp(i \omega t)\Bigg]~. \cr && \label{E5bis}
\end{eqnarray}
In the grating monochromator, a low spectral resolution is
equivalent to a large slit size compared to the spot size of a given
individual frequency at the Fourier plane, and to a small slit size
compared to the spot size of whole spectrum. In this limit, the slit
size does not modify the spatial profile of the output beam, but
modifies spectrum. The output field in this case can be expressed as

\begin{eqnarray}
E_5(x,t) = \mathrm{Re}\Bigg[&&\frac{i}{2\pi} \sqrt{\frac{f
\beta}{f'}} b\left(- \frac{\beta f x}{f'}\right) \cr && \times \int
d \Delta \omega S(\eta \Delta \omega) A(\Delta \omega)
\exp\left(\frac{-i p \Delta \omega f x }{f'}\right) \exp(i \omega
t)\Bigg]~. \label{finalfield}
\end{eqnarray}
The optimal choice for the two-lens system magnification is when
$\beta f/f ' = 1$.  This is the case when the field after the
grating monochromator is perfectly matched to the FEL mode in the
second undulator.

\begin{figure}[tb]
\begin{center}
\includegraphics[width=0.75\textwidth]{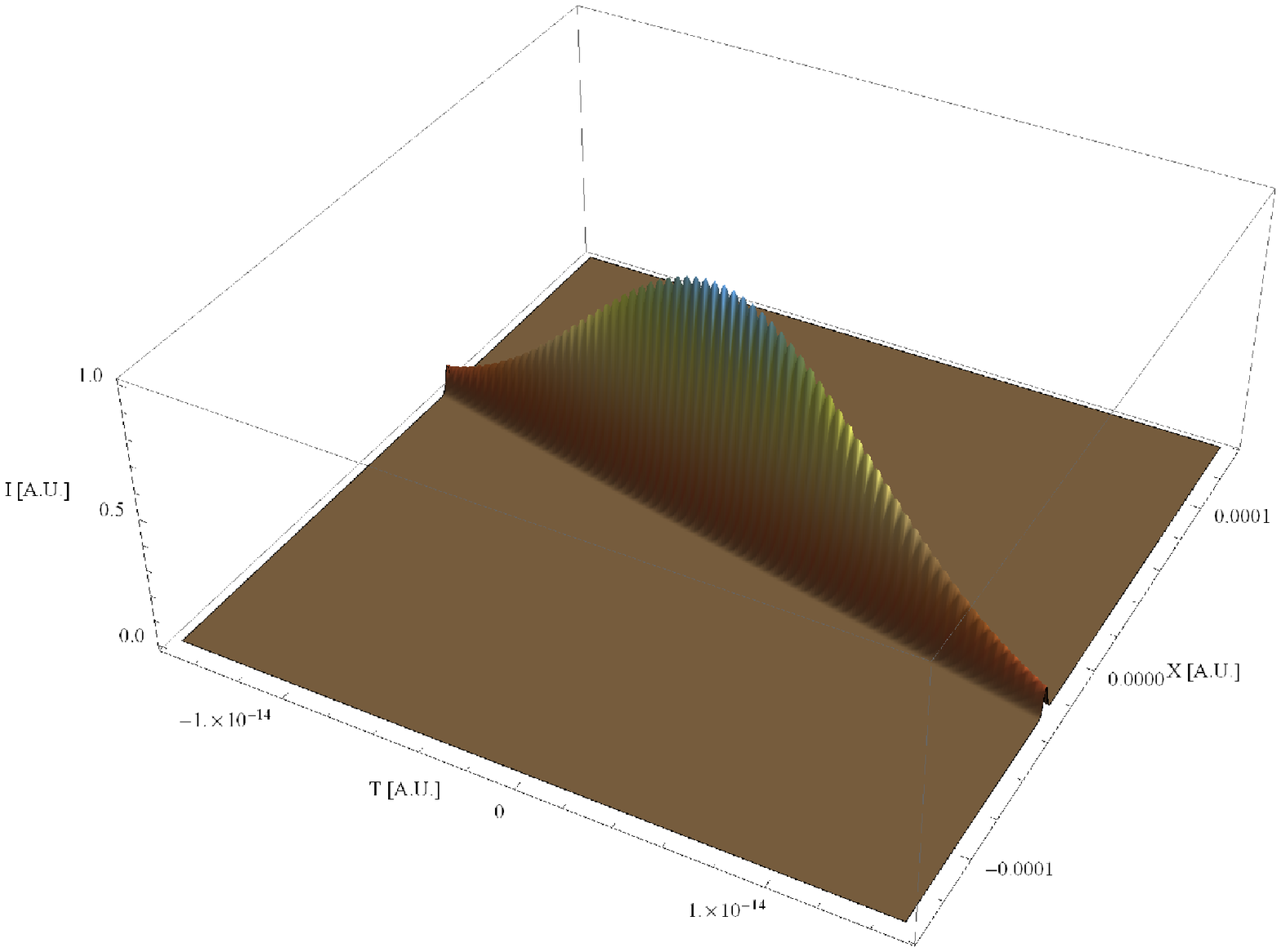}
\end{center}
\caption{Intensity profile in the space-time domain for $\alpha =
100$} \label{3d1}
\end{figure}
\begin{figure}[tb]
\begin{center}
\includegraphics[width=0.75\textwidth]{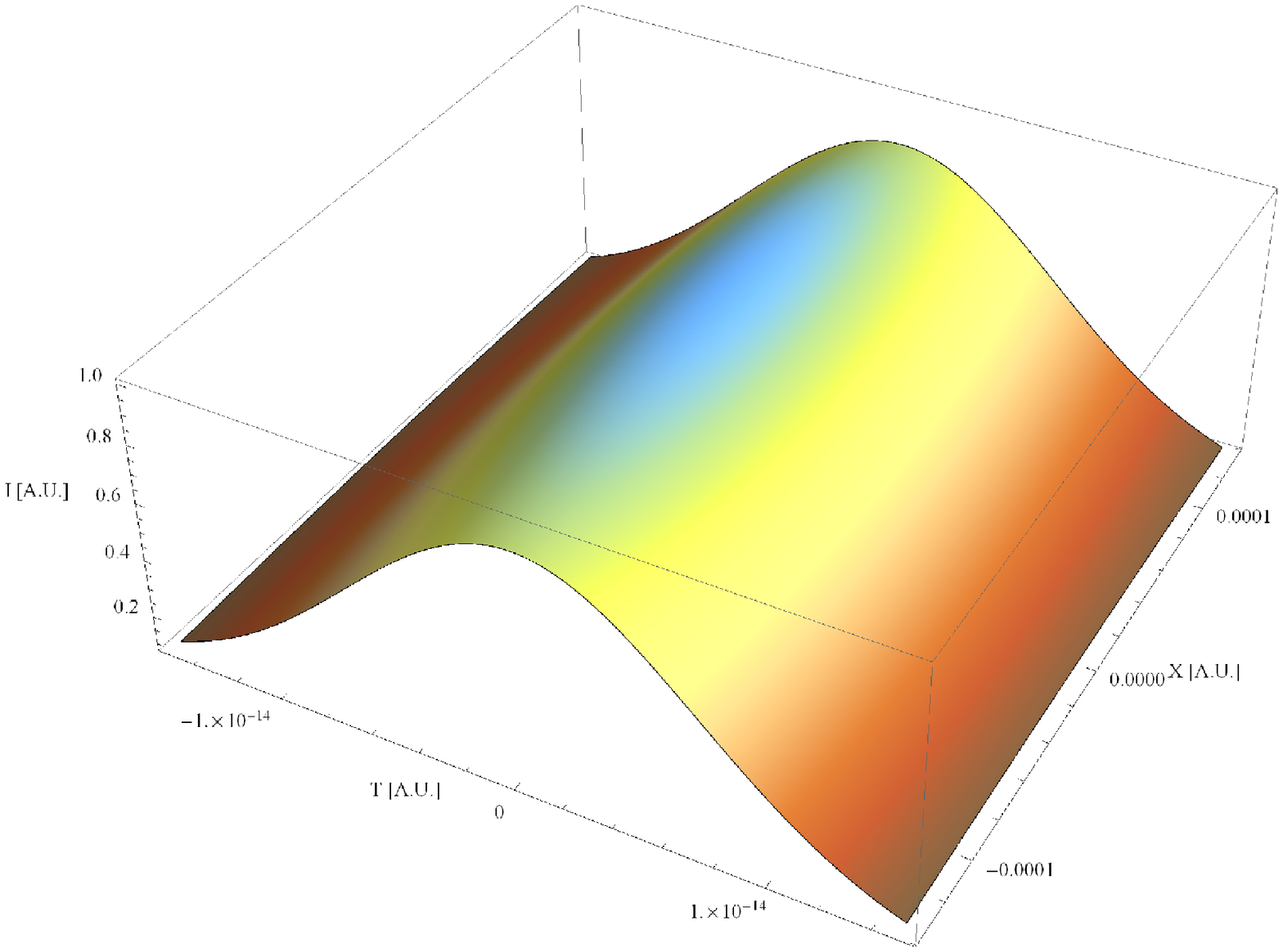}
\end{center}
\caption{Intensity profile in the space-time domain for $\alpha =
0.1$} \label{3d2}
\end{figure}
\begin{figure}[tb]
\begin{center}
\includegraphics[width=0.75\textwidth]{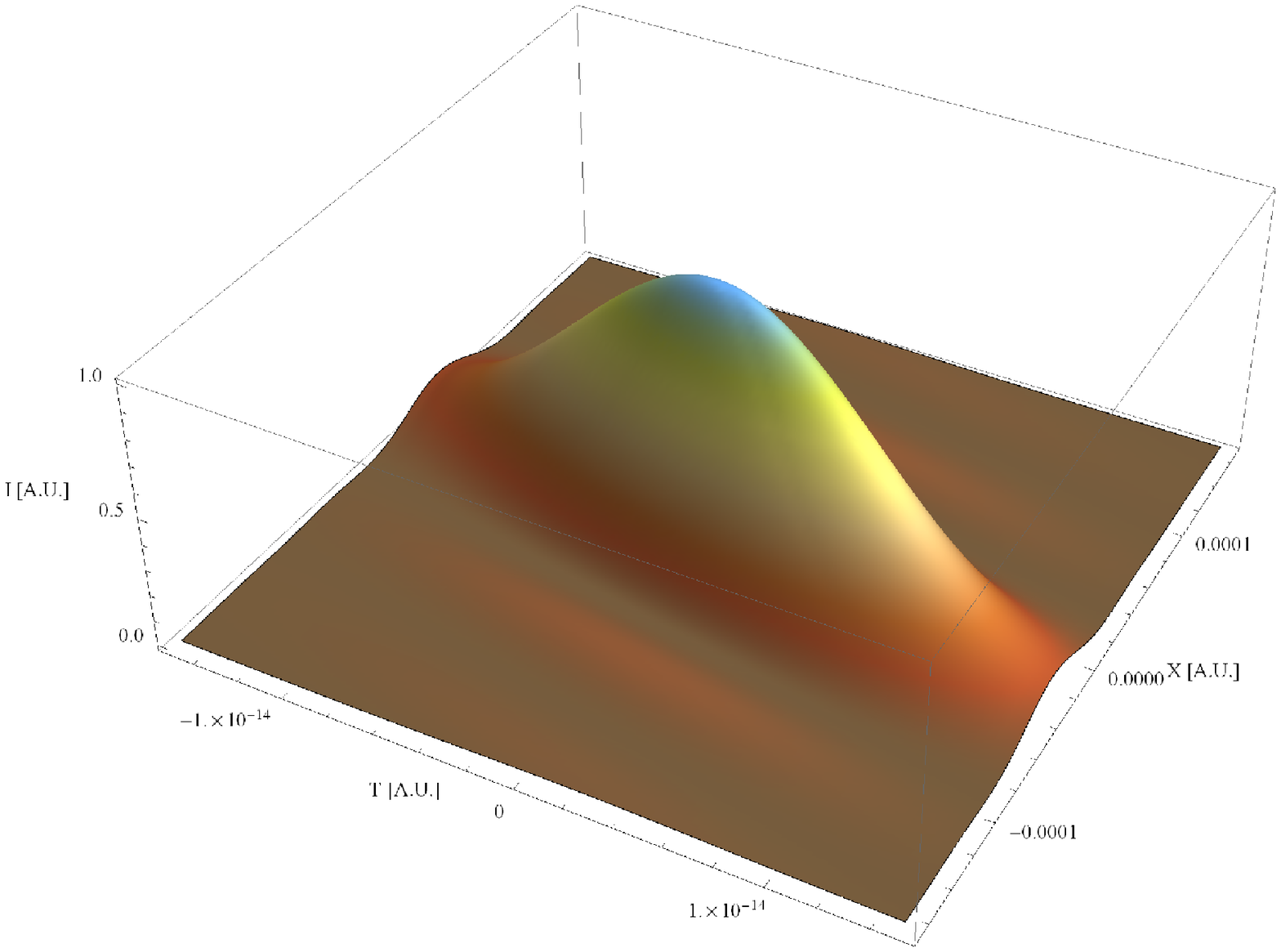}
\end{center}
\caption{Intensity profile in the space-time domain for $\alpha =2$}
\label{3d3}
\end{figure}
\begin{figure}[tb]
\includegraphics[width=1.0\textwidth]{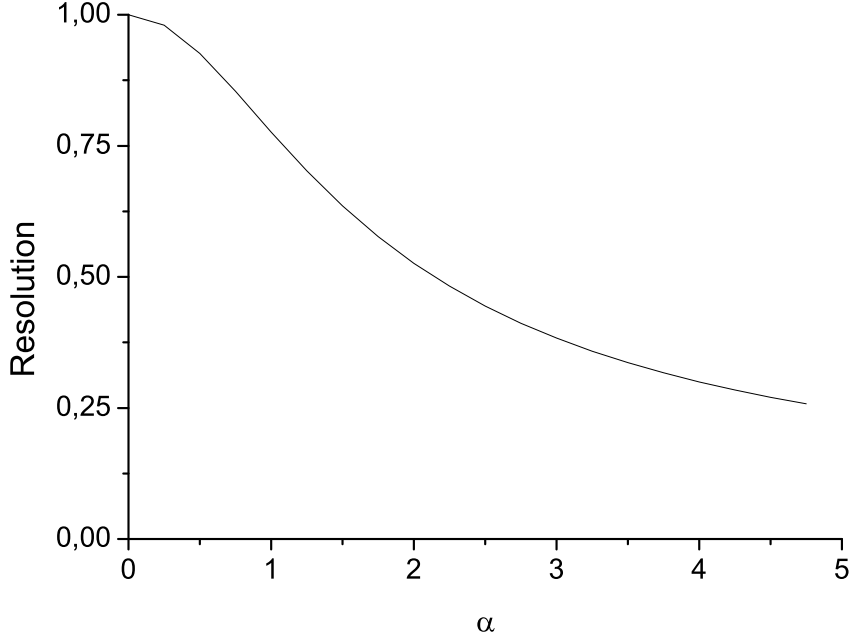}
\caption{Resolving power normalized to the asymptotic case for
$\alpha \ll 1$ as a function of $\alpha$.} \label{Rmerit}
\end{figure}
\begin{figure}[tb]
\includegraphics[width=1.0\textwidth]{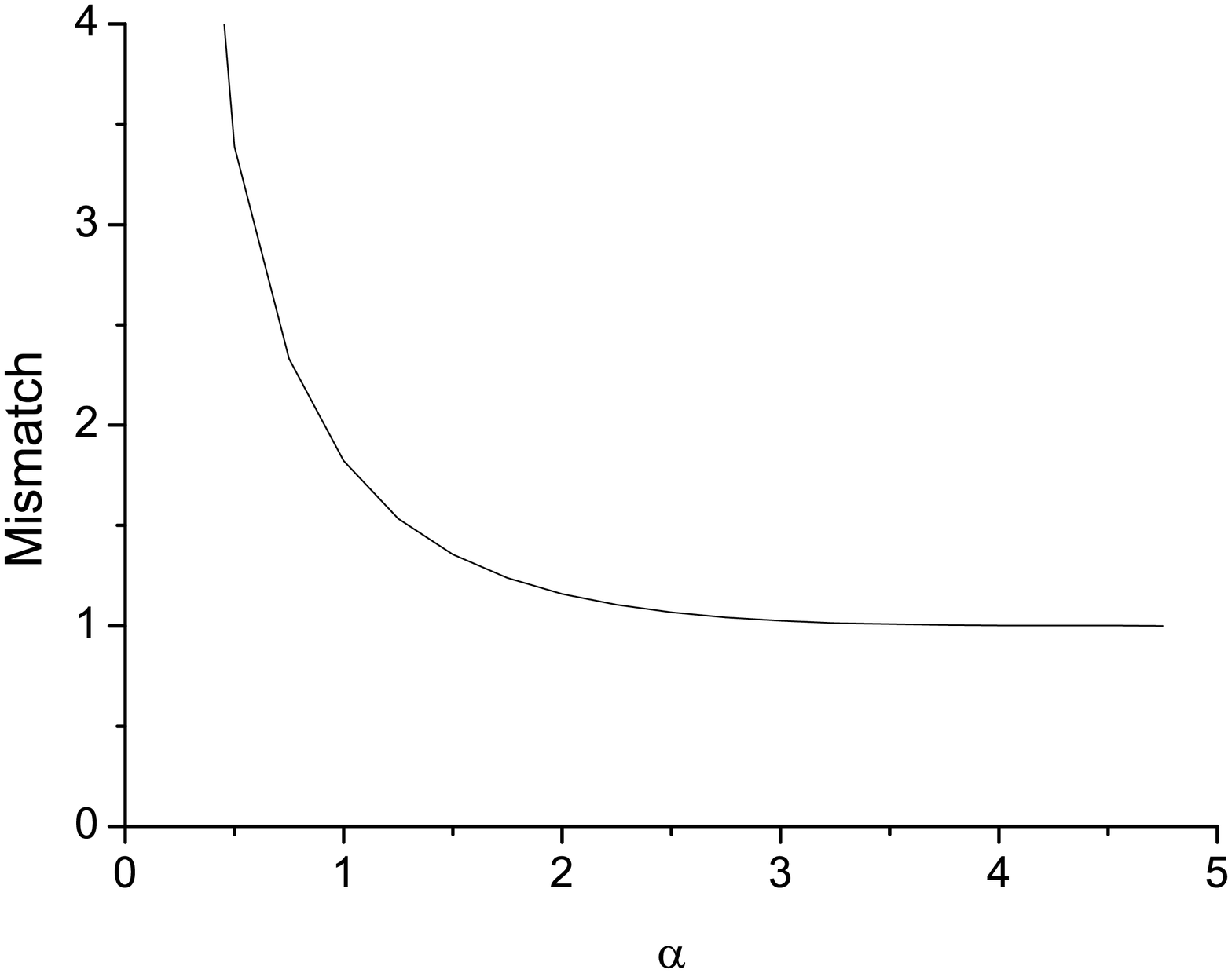}
\caption{Transverse spot size of the photon beam normalized to the
asymptotic case for $\alpha \gg 1$ as a function of $\alpha$. We
assume that the magnification of the two-lens optical system of the
monochromator compensates the astigmatism introduced by the grating,
that is $f \theta_i/f' \theta_D = 1$. } \label{Mmerit}
\end{figure}
We fix the slit function $S$ as

\begin{eqnarray}
&&S(x) = 1 ~~\mathrm{for}~ |x|<d_s \cr && S(x) = 0 ~~\mathrm{for}~
|x|>d_s \label{slit}
\end{eqnarray}
Given a slit with half size $d_s$, we introduce a normalized notion
of slit size

\begin{eqnarray}
\alpha = \frac{d_s}{\sigma_f} = d_s \frac{k \sigma}{\beta
f}\label{alpha}
\end{eqnarray}
We can now plot the intensity profile in the space-time domain for
different values of the normalized slit size $\alpha$. Fig.
\ref{3d1} and Fig. \ref{3d2} qualitatively show the two limiting
situations respectively for $\alpha = 100$ and $\alpha = 0.1$. The
spatiotemporal coupling is evident in Fig. \ref{3d1}. Fig. \ref{3d3}
shows the analogous plot in an intermediate situation, for
$\alpha=2$.

It is possible to show the output characteristics of the radiation
as a function of the slit size by means of universal plots. We first
consider the resolving power $R=(\Delta
\omega/\omega)^{-1}_\mathrm{FWHM}$. We introduce the resolving power
$R_n$ normalized to the inverse of the maximal bandwidth in Eq.
(\ref{Dww}), that is the bandwidth in the limiting case for $\alpha
\ll 1$, as

\begin{eqnarray}
R_n = R \left(\frac{1.18 \theta_i d}{2 \pi \sigma}\right)~.
\label{Rn}
\end{eqnarray}
The behavior of $R_n$ as a function of $\alpha$ is shown in Fig.
\ref{Rmerit}. The resolution of monochromator increases as the slit
size decreases. The $90 \%$ of the maximal resolution level is met
for normalized slit width less than $\alpha < 1$. However, the
energy of the seed pulse decreases proportionally to the decrease of
the slit width. Moreover, decreasing the slit width will also cause
an increase of the output beam size. This will lead to spatial
mismatch between the seed beam and the FEL mode in the second
undulator. The relationship between the beam transverse size (in
terms of FWHM) and slit width is shown in Fig. \ref{Mmerit}, where
we plot the transverse spot size of the photon beam normalized to
the asymptotic case for $\alpha \gg 1$ as a function of $\alpha$. To
summarize, it is not recommended that the normalized slit width be
narrower than unity if a reasonable seed field amplitude is
required.

Finally, a useful figure of merit measuring the spatiotemporal
coupling can be found in \cite{GABO}. Considering the angular
dispersion this parameter can be written as

\begin{eqnarray}
\rho = \int d k_x d \Delta \omega I(k_x, \Delta \omega) \frac{k_x
\Delta \omega}{<(\delta k_x)^2>^{1/2} <(\delta \omega)^2>^{1/2}} ~,
\label{rhopar}
\end{eqnarray}
where

\begin{eqnarray}
&&<(\delta k_x)^2> =    \int d k_x d \Delta \omega I(k_x, \Delta
\omega) k_x^2 ~,\cr &&<(\delta \omega)^2> = \int d k_x d \Delta
\omega I(k_x, \Delta \omega) \Delta \omega^2 ~,\cr && I(k_x, \Delta
\omega) = |E(k_x, \Delta \omega)|^2 \label{extrarho}
\end{eqnarray}
The range of $\rho$ is in $[-1, 1]$ and readily indicate the
severity of these distortions.

\begin{figure}[tb]
\begin{center}
\includegraphics[width=1.0\textwidth]{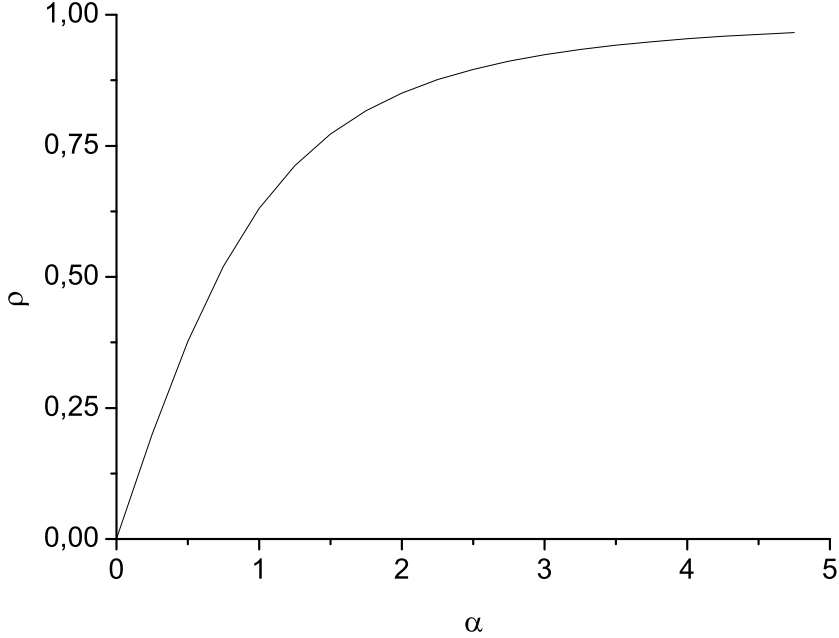}
\end{center}
\caption{Dependence of the spatiotemporal coupling as a function of
$\alpha$ as from Eq. (\ref{rhopar}).} \label{merit2}
\end{figure}
To estimate the pulse front tilt distortion calculate the pulse
front tilt parameter $\rho$ as a function of the slit width
$\alpha$. The results are shown in Fig. \ref{merit2}. It is found to
be larger than $50 \%$ for a slit width $\alpha > 1$. Therefore,
standard tuning of the seed monochromator will lead to significant
spatiotemporal coupling in the seed pulses. The effect of  pulse
front tilt distortion can be reduced if the slit width will be
narrower than $\alpha < 1$. However, the reduction of the pulse
front tilt influence is accompanied by significant loss in seed
signal amplitude.

\section{\label{sec:conc} Conclusions}

To the best of our knowledge, there are no articles reporting on the
impact of pulse front tilt distortions of the seed pulse in the
performance on self-seeding soft X-ray setups. Spatiotemporal
coupling is natural in grating monochromator optics, because the
monochromatization process involves the introduction of angular
dispersion, which is equivalent to pulse front tilt distortion. In
general, it is desirable that the resulting seed pulse be free of
such distortion. This can be achieved only by decreasing the width
of the monochromator slits. On the one hand, decreasing the slit
width increases the resolving power and suppresses the pulse front
tilt distortion. On the other hand, it decreases the seed power and
increases the transverse mismatch with the FEL mode in the second
undulator. As a result, a tradeoff must be reached between
achievable resolution and effective level of the input signal.

Transverse coherence of XFEL radiation is settled without seeding.
This is due to the transverse eigenmode selection mechanism: roughly
speaking, only the ground eigenmode survives at the end of
amplification process. It follows that the spatiotemporal
distortions of the seed pulse do not affect the quality of the
output radiation. They only affect the input signal value.
Therefore, the relevant value for self-seeded operation is the input
coupling factor between the seed pulsed beam and the ground
eigenmode of the FEL amplifier.

In order model the performance of a soft X-ray self-seeded FEL with
a grating monochromator, one naturally starts with the grating
monochromator optical system. One aspect of optimizing the output
characteristics of the self-seeded FEL involves the specification of
spectral width, peak power, pulse-front tilt parameter and
transverse size of the seed pulse as a function of the slit width.
This can be achieved by purely analytical methods. Another aspect of
the problem is the modeling of the FEL process including a seed
pulse with spatio-temporal distortions and transverse mismatching
with the ground FEL eigenmode. This study can be made only with
numerical simulation code.

In this article we restrict our attention to the first part of the
self-seeding process, discussing spatiotemporal coupling of the
electric field in seed pulses. The field amplitude at the exit of
the self-seeding grating monochromator can be obtained using
physical optics rather than a geometrical approach. An analysis of
the behavior of an X-ray optical pulse passing through a grating
monochromator is given in terms of analytical results. In order to
make results useful for practical applications, we have included
numerous universal graphs which can be useful to find a balance
between resolving power, seed power and pulse quality during the
design phase.

\section{Acknowledgements}

We are grateful to Massimo Altarelli, Reinhard Brinkmann,
Serguei Molodtsov and Edgar Weckert for their support and their interest during the compilation of this work.

\end{document}